\documentclass[aps,pra,superscriptaddress,amsmath,amssymb,preprintnumbers,showpacs, twocolumn]{revtex4}

\usepackage{amssymb}
\usepackage{graphicx}
\usepackage{subfigure}
\usepackage{bm}
\usepackage{color}

\newcommand{\Ket}[1]{\left\vert #1\right\rangle}
\newcommand{\Bra}[1]{\left\langle #1\right\vert}

\newcommand{\BraKet}[2]{\left\langle#1\vert #2\right\rangle}
\newcommand{\KetBra}[2]{\left\vert#1\right\rangle\left\langle#2\right\vert}
\newcommand{\Projector}[1]{\KetBra{#1}{#1}}

\renewcommand{\eqref}[1]{(\ref{#1})} 

\newcommand{\Change}[2]{#2}

\begin{document}

\title{Quantum Zeno Subspaces induced by Temperature}

\author{B. Militello}
\affiliation{Dipartimento di Fisica dell'Universit\`{a} di
Palermo, Via Archirafi 36, 90123 Palermo, Italy}
\email{bdmilite@fisica.unipa.it}

\author{M. Scala}
\affiliation{Dipartimento di Fisica dell'Universit\`{a} di
Palermo, Via Archirafi 36, 90123 Palermo, Italy}

\author{A. Messina}
\affiliation{Dipartimento di Fisica dell'Universit\`{a} di
Palermo, Via Archirafi 36, 90123 Palermo, Italy}

\begin{abstract}
We discuss the partitioning of the Hilbert space of a quantum
system induced by the interaction with another system at thermal
equilibrium, showing that the higher the temperature the more
effective is the formation of Zeno subspaces. We show that our
analysis keeps its validity even in the case of interaction with a
bosonic reservoir, provided appropriate limitations of the
relevant bandwidth.
\end{abstract}

\maketitle



\section{Introduction}

The quantum Zeno effect (QZE), in its original form, is the
inhibition of the dynamics of a physical system due to frequent
measurements of its state~\cite{ref:MishraSudarshan}. In fact, the
combined action of unitary evolutions for a short time and
projection operators gives rise to an effective quantum evolution
which becomes closer and closer to a simple projection on the
initial state of the system, when the time interval between two
measurements becomes shorter and shorter. This effect has been
demonstrated in various physical systems~\cite{ref:Itano1990,
ref:Fischer1997}.

Subsequent studies have raised the problem of analyzing the
inhibition of the dynamics induced by continuous measurements,
meant as dissipative processes~\cite{ref:Schulman1998}. Indeed,
the emission of radiation from a quantum system could be thought
of as measurement process, since it says to the observer in which
state the system was before the emission. Apart from philosophical
discussions, the presence of a strong decay can be responsible for
a dynamical decoupling, hindering the dynamics induced by other
couplings~\cite{ref:Panov1999a, ref:Panov1999b, ref:Militello2011,
ref:Scala2010}.

Further developments has brought to the idea that even a unitary
coupling can be responsible for the inhibition of the dynamics
induced by smaller couplings~\cite{ref:PascazioFacchi2001,
ref:Militello2001A, ref:Militello2001B}, eventually leading to the
concept of Zeno subspaces, i.e., the formation of suitable
invariant subspaces that cancels the effects of weak couplings, as
shown by Facchi and Pascazio~\cite{ref:PascazioFacchi2002}. The
partitioning of the Hilbert space can be obtained also by \lq
bang-bang\rq\, (BB) decoupling. The theories of QZE and BB have
been recently unified~\cite{ref:PascazioFacchi2004,
ref:PascazioFacchi2008, ref:PascazioFacchi2010}. It is worth
mentioning that A. Peres has shown how similar behaviors can be
observed even in the frame of classical theory~\cite{ref:Peres1}.

The importance of QZE is related to conceptual aspects and to the
foundation of quantum mechanics~\cite{ref:Home1997}, but it is
also witnessed by a variety of applications in the fields of
quantum information and nanotechnologies~\cite{ref:App1, ref:App2,
ref:App3}.

About a decade ago, Ruseckas has studied the influence of the
temperature of the detector on the quantum Zeno effect, showing
that a higher temperature of the measurement apparatus can enhance
the inhibition of the dynamics at short time, then resulting in a
more evident Zeno effect by pulsed
measurements~\cite{ref:Ruseckas2002}. More recently, Maniscalco
{\it et al} have analyzed the crossover from quantum Zeno effect
to Inverse Zeno effect (IZE, meant as the enhancement of the
dynamics due to repeated
measurements~\cite{ref:PascazioFacchi2001}) in the Quantum
Brownian motion, also bringing to light a certain role of
temperature in the occurrence of such
effects~\cite{ref:Sabrina2006}. Very recently, Bhaktavatsala and
Kurizki have analyzed the influence of the statistics of the
environment (fermions vs bosons) to the QZE-IZE
crossover~\cite{ref:Bhaktavatsala2011}.

In this paper we study the role of temperature in the partitioning
of the Hilbert space of a physical system, i.e. we analyze the
formation of Zeno subspaces due to the thermal bosons a system is
interacting with. Starting from the study --- analytical and
numerical --- of the interaction between a three-state system and
a single harmonic oscillator at thermal equilibrium, and between
the same three-state system and a finite number of thermalized
harmonic oscillators, we find a universal feature corresponding to
the partitioning of the Hilbert space of the three-state system at
high temperatures. Our analysis allows us to single out a critical
temperature for the observation of such a partitioning. The
scaling of this threshold temperature easily allows a
generalization to the case of infinite harmonic oscillators.

In the next section we show the basic of this effect by studying
the single harmonic oscillator case. In the third section, we
extend the discussion to the case of a finite but arbitrary number
of harmonic oscillators, and in the fourth section we give an
estimation of a temperature high enough to observe the formation
of Zeno subspaces.  Finally, in the last section we provide a
detailed discussion of the results.

\section{Single Harmonic Oscillator}

We consider a three-state system coupled with an harmonic
oscillator tuned close to the transition $2\rightarrow 3$.

The relevant Hamiltonian is ($\hbar=1$):
\begin{eqnarray}
  \nonumber
  H &=& \sum_{k=1,2,3}\omega_k\Projector{k}+\Omega
  (\KetBra{1}{2}+\KetBra{2}{1})\\ &+& \omega \hat{a}^\dag\hat{a} +
  g(\hat{a}\KetBra{2}{3}+\hat{a}^\dag\KetBra{3}{2})\,,
  \label{eq:SingleOcillatorHamiltonian}
\end{eqnarray}
where $\Ket{k}$ with $k=1,2,3$ denotes the generic state of the
three-state system, $\omega_k$ is the relevant energy, $\Omega$
gives the strength of the coupling between $1$ and $2$, $\omega$
is the frequency of the harmonic oscillator (whose annihilation
and creation operators are $\hat{a}$ and $\hat{a}^\dag$) and $g$
is the coupling strength between the three-state system and the
oscillator.

For the sake of simplicity we are assuming that the coupling
constant $g$ is real, but all the results we shall show are valid
also in the most general case where $g$ is complex and the
coupling is $g\hat{a}\KetBra{2}{3}+g^*\hat{a}^\dag\KetBra{3}{2}$
(see the discussion in the end of Appendix
\ref{App:Many_Oscillators}).

This Hamiltonian is structured as a set of invariant blocks of the
form:
\begin{equation}
  H_n = \left(
  \begin{array}{ccc}
    \omega_1 + n\omega & \Omega & 0 \cr
    \Omega & \omega_2 + n\omega & g\sqrt{n+1} \cr
    0 & g\sqrt{n+1} & \omega_3 + (n+1)\omega
  \end{array}
  \right)\,,
  \label{eq:SingleOcillatorBlock}
\end{equation}
corresponding to the triplet $\Ket{1}\Ket{n}$, $\Ket{2}\Ket{n}$,
$\Ket{3}\Ket{n+1}$.

The relevant evolution operator is $U_n(t)=\exp(-i H_n t)$. When
the condition $g\sqrt{n+1}\gg\Omega$ is fulfilled, such unitary
operator does not significantly change the population of the state
$\Ket{1}\Ket{n}$. Qualitatively, we can say that in such case the
coupling $\Omega (\Ket{1}\Bra{2}+\Ket{2}\Bra{1})$ is a small
perturbation which does not affect much the dynamics of the state
$\Ket{1}\Ket{n}$ which is an eigenstate in the case $\Omega=0$.

Suppose now that the two subsystems are initially uncorrelated and
that the harmonic oscillator is at thermal equilibrium:
\begin{equation}
  \rho = \rho_A \otimes \rho_B\,,
\end{equation}
with
\begin{equation}
  \rho_A = \Projector{1}\,,
\end{equation}
\begin{equation}
  \rho_B = \sum_n p_n
  \Projector{n}\,,   \qquad  p_n=Z^{-1} \exp\left(-\frac{n\omega}{k_B T}\right)\,,
\end{equation}
where $Z=[1-\exp(-\omega/(k_B T))]^{-1}$.

The state at time $t$ can be written as:
\begin{equation}
  \rho(t) = \sum_n p_n U_n(t) \Ket{n}\Ket{1}\Bra{1}\Bra{n} U_n^\dag(t) \,,
\end{equation}
so that the survival probability of the atomic state $\Ket{1}$ is
given by:
\begin{equation}
  P(t) = \sum_n p_n \left|\Bra{1}\Bra{n} U_n(t) \Ket{n}\Ket{1}\right|^2\,.
\end{equation}

We shall rigorously prove that for any $\varepsilon > 0$ it is
possible to find a temperature $T_\varepsilon$ such that for $T >
T_\varepsilon$ it is $P(t) > 1-\varepsilon$ at every time. In the
Appendix \ref{App:Many_Oscillators} we prove that for any
$0<\varepsilon<1$ it is possible to find an index $n_\varepsilon$
such that for $n\ge n_\varepsilon$ it turns out
$\left|\Bra{1}\Bra{n} U_n(t)
\Ket{n}\Ket{1}\right|^2\ge\sqrt{1-\varepsilon}$ for every $t$
(consider the special case $D=1$, with $c_1=g\sqrt{n+1}$).
Therefore, the survival probability satisfies the following
relations:
\begin{eqnarray}
\nonumber
  P(t) &\ge& \sum_{n<n_\varepsilon} p_n \left|\Bra{1}\Bra{n} U_n(t) \Ket{n}\Ket{1}\right|^2\\
\nonumber
       &+& \sum_{n\ge n_\varepsilon} p_n \sqrt{1-\varepsilon}\,\,
       \ge \,\, \sum_{n\ge n_\varepsilon} p_n \sqrt{1-\varepsilon}\\
       &=& \exp\left(-\frac{\omega n_\varepsilon}{k_B T}\right) \sqrt{1-\varepsilon}\,,
\end{eqnarray}
where we have used the relation $\sum_{n\ge n_\varepsilon} x^n = x^{n_\varepsilon}/(1-x)$.

Let us now consider a temperature $T > T_\varepsilon = -(2\omega
n_\varepsilon)/(k_B \log(1-\varepsilon))$, for which one has
$\exp(-\omega n_\varepsilon/(k_B
T_\varepsilon))>\sqrt{1-\varepsilon}$. Then, for such high
temperatures one has $P(t) > 1-\varepsilon$ for any $t$. This is a
clear manifestation of the formation of temperature-induced Zeno
subspaces. Indeed, the higher the temperature, the more the
coupling between atomic states $\Ket{1}$ and $\Ket{2}$ is
neutralized by the coupling between $\Ket{2}$ and $\Ket{3}$
mediated by the harmonic oscillator.

It is important to note that when $T\approx 0$ the only block
really involved in the dynamics is the one corresponding to $n=0$,
and the inhibition of the time evolution can occur only if $g\gg
\Omega$. On the contrary, when the temperature increases, there
are a number of blocks effectively involved in the dynamics, and
in most of them it happens that the coupling constant
$g\sqrt{n+1}$ exceeds $\Omega$, even if $g$ itself is not large.
In the $T\rightarrow\infty$ limit all the blocks (an infinite
number) are involved and in the majority of them the condition
$g\sqrt{n+1}\gg\Omega$ is fulfilled.

\begin{figure}
\includegraphics[width=0.45\textwidth, angle=0]{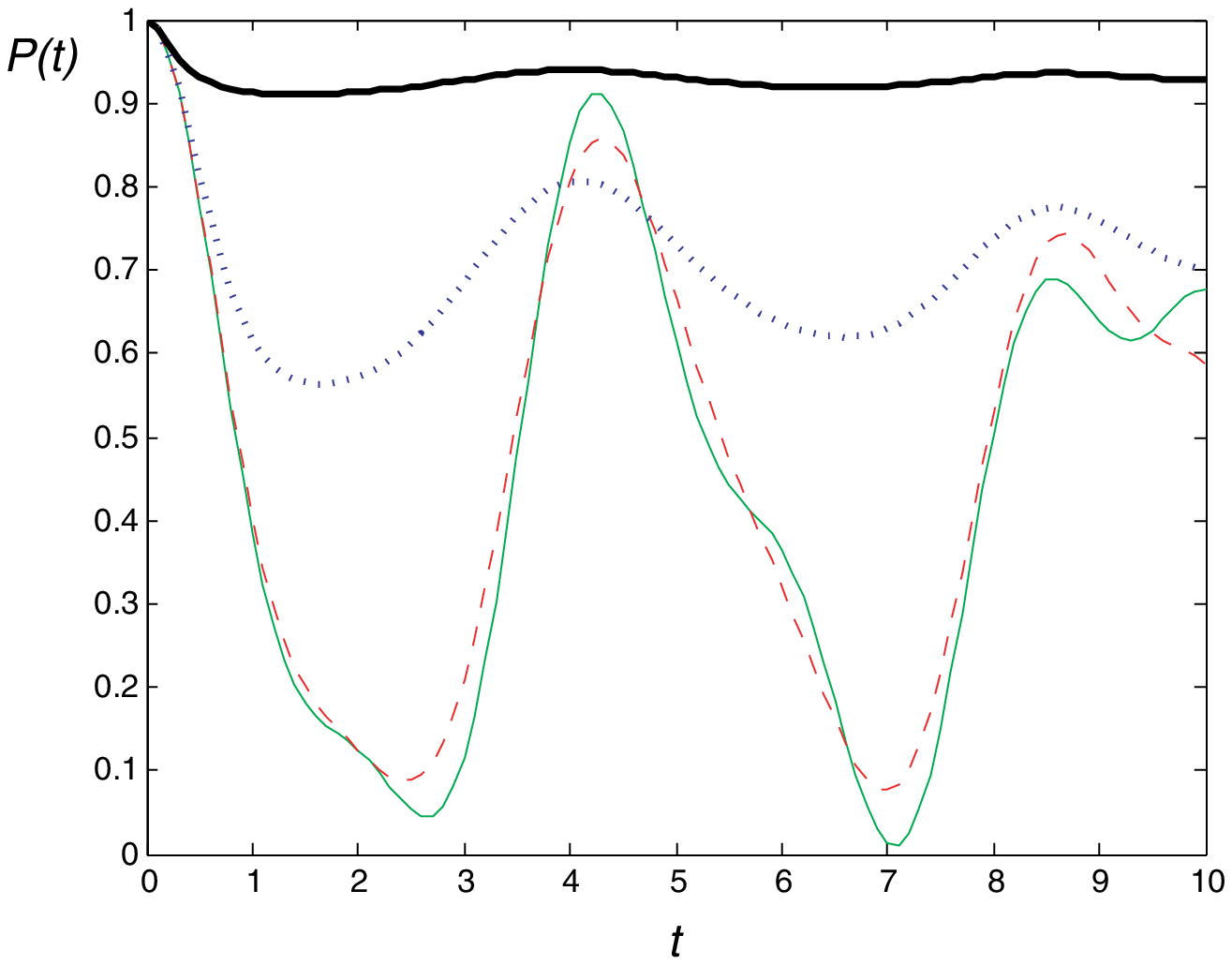}
\caption{(Color online) The survival probability of the atomic
state $\Ket{1}$ as a function of time (in units of $\Omega^{-1}$),
at different temperatures: $k_B T/\omega_{23}=0.1$ (green solid
line), $k_B T/\omega_{23}=1$ (red dashed line), $k_B
T/\omega_{23}=10$ (blue dotted line), $k_B T/\omega_{23}=100$
(black bold line). Here $\omega_1/\Omega=20$,
$\omega_2/\Omega=19$, $\omega_3=0$,
$\omega_{23}=\omega_2-\omega_3$, $D=1$, $g/\Omega=1$. The harmonic
oscillator is tuned to the $2\rightleftarrows 3$ transition.}
\label{fig:Single_Oscillator}
\end{figure}

To support our previous analysis, we show in
Fig.~\ref{fig:Single_Oscillator} the evolution of the survival
probability of the state $\Ket{1}$ when the system is interacting
with an oscillator resonant to the transition $2 \rightarrow 3$,
for different temperatures. It is well visible that as the
temperature increases the survival probability tends toward unity
at every time.

\section{Many Harmonic Oscillators}

The previous result can be generalized to a set of $D$ harmonic
oscillators interacting with a three-state system. The relevant
Hamiltonian reads:
\begin{eqnarray}
  \nonumber
  H &=& \sum_{k=1,2,3}\omega_k\Projector{k}+\Omega
  (\KetBra{1}{2}+\KetBra{2}{1})\\ %
  \nonumber
  &+& \sum_{k=1}^D\tilde{\omega}_k \hat{a}_k^\dag\hat{a}_k +
  \sum_{k=1}^D
  g_k(\hat{a}_k\KetBra{2}{3}+\hat{a}_k^\dag\KetBra{3}{2})\,.\\
  \label{eq:ManyOcillatorsHamiltonian}
\end{eqnarray}

Each set $\{\Ket{1}\Ket{n_1,n_2,...,n_D}$,
$\Ket{2}\Ket{n_1,n_2,...,n_D}$, $\Ket{3}\Ket{n_1+1,n_2,...,n_D}$,
...$\Ket{3}\Ket{n_1,n_2,...,n_D+1}\}$ individualizes an invariant
subspace, corresponding to the following invariant block:
\begin{widetext}
\begin{equation}
  H_{n_1 ... n_D} = \left(
  \begin{array}{ccccc}
    \omega_1 + \delta & \Omega & 0 & ... & 0 \cr
    \Omega & \omega_2 + \delta & g_1\sqrt{n_1+1} & ... &  g_D\sqrt{n_D+1} \cr
    0 & g_1\sqrt{n_1+1} & \omega_3 + \delta + \tilde{\omega}_1 & ... & 0 \cr
    \vdots  & \vdots & \vdots & \ddots & \vdots \cr
    0 & g_D\sqrt{n_D+1} & 0 & ... & \omega_3 + \delta + \tilde{\omega}_D \cr
  \end{array}
  \right)\,,
  \label{eq:ManbyOcillatorsBlock}
\end{equation}
\end{widetext}
with $\delta=\sum_k n_k\tilde{\omega}_k$. The only non vanishing
terms of this matrix belong to the second row, to the second
column (all the couplings involve state $\Ket{2}\Ket{n_1,...n_D}$)
and to the diagonal (free energies). Similarly to the case of one
harmonic oscillator, when it happens that
$g_k\sqrt{n_k+1}\gg\Omega$\, $\forall k$, one can assume that the
Hamiltonian in \eqref{eq:ManbyOcillatorsBlock} with $\Omega=0$ is
the \lq unperturbed\rq\, Hamiltonian, which has
$\Ket{1}\Ket{n_1,...n_D}$ as eigenstate; then one can consider the
effects of the perturbation (the coupling
$\Omega(\Ket{1}\Bra{2}+\Ket{2}\Bra{1})$), which does not change
much the previous situation, leaving the state
$\Ket{1}\Ket{n_1,...,n_D}$ as an eigenstate up to terms of the
order $\Omega/(g_k\sqrt{n_k+1})$. This implies that for very large
$n_1$, $n_2$, ... $n_D$ the population of the state
$\Ket{1}\Ket{n_1,...n_D}$ does not significantly change.

The circumstance that the state $\Ket{1}\Ket{n_1,...,n_D}$ is
close to an eigenstate of the Hamiltonian for large excitation
numbers is rigorously proven in the Appendix
\ref{App:Many_Oscillators}.

Similarly to the case of a single harmonic oscillator, for large
enough temperature the blocks with large excitation numbers are
more and more involved in the dynamics and the subspace separation
becomes more and more significant.

\begin{figure}
\includegraphics[width=0.45\textwidth, angle=0]{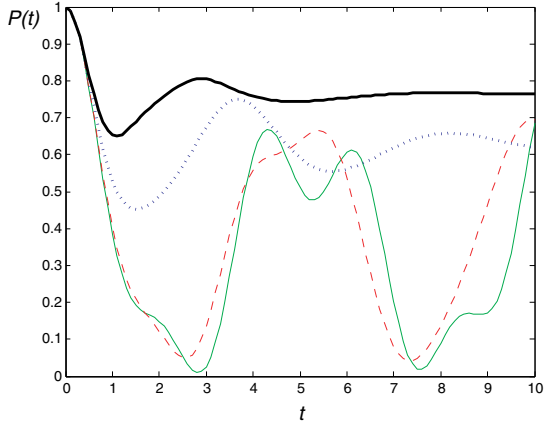}
\caption{(Color online) The survival probability of the atomic
state $\Ket{1}$ as a function of time (in units of $\Omega^{-1}$),
at different temperatures: $k_B T/\omega_{23}=0.1$ (green solid
line), $k_B T/\omega_{23}=1$ (red dashed line), $k_B
T/\omega_{23}=5$ (blue dotted line), $k_B T/\omega_{23}=10$ (black
bold line). Here $\omega_1/\Omega=20$, $\omega_2/\Omega=19$,
$\omega_3=0$, $\omega_{23}=\omega_2-\omega_3$, $D=4$,
$g_k/\Omega=g/\Omega=0.5$. The four oscillators are tuned to the
frequencies: $\tilde{\omega}_1 = \omega_{23}$, $\tilde{\omega}_2
\approx 0.996\,\omega_{23}$, $\tilde{\omega}_3 \approx
0.992\,\omega_{23}$, $\tilde{\omega}_4 \approx
0.987\,\omega_{23}$.} \label{fig:Many_Oscillators}
\end{figure}

In Fig.~\ref{fig:Many_Oscillators} we show the time evolution of
the survival probability of the state $\Ket{1}$ when the
three-state system is interacting with a finite number of
oscillators having frequencies close to the $2 \rightarrow 3$
transition. In these calculations we have taken $g=0.5$ and $D=4$,
in order to keep constant the quantity $g^2 D$, which resembles
the zero-temperature decay rate in a bath $\Gamma=g^2(\omega)
D(\omega)$, where $D(\omega)$ is the density of modes of frequency
$\omega$ \cite{ref:Gardiner, ref:Petruccione}. In order to better
afford numerical calculations in the presence of a larger Hilbert
space, we have considered values of the temperature lower than
those considered in the case $D=1$. The Zeno-like effect is still
appreciable anyway. Indeed, similarly to the case of a single
harmonic oscillator, the trend of the survival probability as
temperature increases is well visible.

\section{Threshold Temperature}

Let us now study in detail the dependence of the appearance of
Zeno subspaces on the temperature.

Generally speaking, if the initial state of a quantum system has
an overlap $\sqrt{\chi} \, \mathrm{e}^{i\varphi}$ with an
eigenstate of the Hamiltonian, such that $\chi>1/2$, then the
survival probability can never be smaller than $(\chi -
(1-\chi))^2 = (2\chi-1)^2$. We prove this statement in the
Appendix \ref{App:SurvivalProbability}

From Appendix \ref{App:Many_Oscillators} we know that in a given
block characterized by $n_1, n_2,...,n_D$ with $c\equiv \sum_l
g_l^2 n_l$ large enough, there exists an eigenstate, say
$\Ket{\Psi_{n1,...,n_D}}$, whose overlap with the state
$\Ket{A_{n_1, n_2,...,n_D}}\equiv\Ket{1}\Ket{n_1, n_2,...,n_D}$
is:
\begin{eqnarray}
\nonumber
\left|\BraKet{\Psi_{n1,...,nD}}{A_{n_1, n_2,...,n_D}}\right| && \\
&&\!\!\!\!\!\!\!\!\!\!\!\!\!\!\!\!\!\!\!\!\!\!\!%
\!\!\!\!\!\!\!\!\!\!\!\!\!\!\!\!\!\!\!\!\!\!\!\!%
\!\!\!\!\!\!\!\!\!\!\!\!\!\!\!\!\!\!\!\!\!\!\!\!%
\ge \, %
\frac{m c^2}{(m^2 c^4 + 16 M^2 \Omega^2 c^2 + 4 m^2 M^2
\Omega^2)^{1/2}}\,,
\end{eqnarray}
where $m = \min_j(|\omega_3-\omega_1+\tilde{\omega}_j|)$ and $M =
\max_j(|\omega_3-\omega_1+\tilde{\omega}_j|)$.

To obtain the previous formula, according to our analysis in
Appendix \ref{App:Many_Oscillators}, we require either
$\tilde{\omega}_j > \omega_1-\omega_3$ \, $\forall k$ or
$\tilde{\omega}_j < \omega_1-\omega_3$ \, $\forall k$, and $m>0$
anyway. For
\begin{equation}
c \ge c_\varepsilon = \frac{4 M \Omega}{m} \,
\sqrt{\frac{1+(1-\varepsilon)^{1/4}}{1-(1-\varepsilon)^{1/4}}}
\approx \frac{8 \sqrt{2}\, \Omega M}{m} \,
\varepsilon^{-\frac{1}{2}} \,,
\end{equation}
the squared modulus of the overlap is larger than
$\chi=((1-\varepsilon)^{1/4}+1)/2$, so that the survival
probability of the state $\Ket{A_{n_1, n_2,...,n_D}}$ can never be
smaller than $\sqrt{1-\varepsilon}$.

Let us consider the part of the bosonic Hilbert space made of
those invariant subspaces where it happens that the relevant $c$
is smaller than $c_\varepsilon$, and introduce the relevant
partition function:
\begin{eqnarray}
\nonumber
Z_\varepsilon &\equiv& Z(\sum_j g_j^2 n_j < c_\varepsilon^2)\\
&=&  \sum_{\sum_k g_k^2 n_k <\, c_\varepsilon^2}
\mathrm{e}^{-\sum_k \tilde{\omega}_k n_k/(k_B T)} \,.
\end{eqnarray}
The smaller $Z_\varepsilon$ the more the atomic state $\Ket{1}$
does not evolve, and it turns out that such a state is essentially
separated from the other two atomic states.

In the following we will show that the contribution of
$Z_\varepsilon$ to the total partition function becomes negligible
in the limit $T\rightarrow\infty$.

{\it Calculation on the hypercube} --- The condition $c \ge
c_\varepsilon$ is surely guaranteed if there is at least one boson
number $n_k \ge n_\varepsilon$, where $n_\varepsilon$ is the
smallest integer number larger than $c_\varepsilon^2 /
g_{\mathrm{min}}^2$ with $g_{\mathrm{min}}=\min_j |g_j|$.
Therefore, the condition $c \ge c_\varepsilon$ is {\it a fortiori}
satisfied if $n_k \ge n_\varepsilon$ \, $\forall k$.

The ratio between $Z_\varepsilon$ and the total partition function
is:
\begin{equation}
  \frac{Z_\varepsilon}{Z_\mathrm{tot}} \, \le \, \frac{Z(\{n_j < n_\varepsilon\})}{Z_{\mathrm{tot}}} = \prod_{j=1}^D (1-x_j^{n_\varepsilon})\,,
\end{equation}
where $Z(\{n_j < n_\varepsilon\})$ is meant as the partition
function corresponding to $n_j < n_\varepsilon$ \, $\forall j$:
\begin{eqnarray}
\nonumber Z(\{n_j < n_\varepsilon\}) &=& \prod_{j=1}^D \,\left(
\sum_{n_j<n_\varepsilon} \mathrm{e}^{-\tilde{\omega}_j n_j / (k_B
T)}\right)\\
&=& \prod_{j=1}^D \,\frac{1-x_j^{n_\varepsilon}}{1-x_j}\,,
\end{eqnarray}
while
\begin{equation}
  Z_{\mathrm{tot}} = \prod_{j=1}^D (1-x_j) \,,
\end{equation}
with $x_j=\exp(-\tilde{\omega}_j/(k_B T))$. Introducing
$\tilde{\omega}_{\mathrm{max}}=\max_j \tilde{\omega}_j$ and $x
\equiv \min_j x_j = \exp(-\tilde{\omega}_{\mathrm{max}}/(k_B T))$,
one finds
\begin{equation}
  \frac{Z_\varepsilon}{Z_{\mathrm{tot}}} \, \le \, (1-x^{n_\varepsilon})^D\,.
\end{equation}
The request $Z_\varepsilon\,/Z_{\mathrm{tot}} <
1-\sqrt{1-\varepsilon}\equiv \alpha_\varepsilon$, translates into
\begin{equation}
  T > -\frac{\tilde{\omega}_{\mathrm{max}} n_\varepsilon}{k_B\,\log(1-\alpha_\varepsilon^{1/D})} \,.
\end{equation}

In the limit of small $\varepsilon$ we have that $c_\varepsilon$
scales as $8 \sqrt{2} \, \Omega M m^{-1}\varepsilon^{-1/2}$ and
that $\log(1-\alpha_\varepsilon^{1/D}) \approx
-(\varepsilon/2)^{1/D}$ provided $D$ is not too large (we need
$\alpha_\varepsilon^{1/D} \ll 1$). Therefore, under such
hypotheses, it is sufficient to have:
\begin{equation}
  T > T_\mathrm{c}(\varepsilon) \approx \, \frac{2^6 \, M^2 \Omega^2\,\tilde{\omega}_{\mathrm{max}}}{m^2 \, g_{\mathrm{min}}^2\,k_B} \,
  \left(\frac{2}{\varepsilon}\right)^{\frac{D+1}{D}}\, D
  \,,
\end{equation}
which eventually guarantees that the survival probability of the
state $\Ket{1}$ is always larger than $1-\varepsilon$.

{\it Calculation on the hypersphere} --- Let us now calculate the
$Z_\varepsilon$ partition function using the approximation to the
continuum, which better applies for high temperature (so that the
Boltzmann factors do not change much with respect to the $n_j$'s):
\begin{eqnarray}
  \nonumber
  Z_\varepsilon &=& \, \sum_{\sum_k g_k^2 n_k <\, c_\varepsilon^2} \mathrm{e}^{-\sum_j \tilde{\omega}_j n_j/(k_B T)} \\
  \nonumber
  &\approx& \, \int ... \int_{\sum_n y_n^2 <\, c_\varepsilon^2} \mathrm{e}^{-\sum_j \tilde{\omega}_j n_j /(k_B T)} \prod_{n=1}^D \frac{2 y_n
  \mathrm{d}y_n}{g_n^2}\,,\\
  \label{eq:PartFuncSphereEpsilon}
\end{eqnarray}
where we have introduced the variables $y_k = g_k\sqrt{n_k}$, so
that $\mathrm{d}n_k = 2 g_k^{-2} y_k \mathrm{d}y_k$.

In the large temperature limit the exponential in
Eq.~(\ref{eq:PartFuncSphereEpsilon}) approaches unity and it turns
out to be:
\begin{equation}\label{eq:PartFuncSphereEpsilonExplicit}
  \frac{Z_\varepsilon}{Z_{\mathrm{tot}}} \approx \frac{2^D \, G_D \, c_\varepsilon^{2D}}{\prod_n g_n^2}
  \frac{\prod_n \tilde{\omega}_n}{(k_B T)^D}
  \,,
\end{equation}
where $G_D=(2^D D!)^{-1}$ is the appropriate geometric factor (see
the appendix \ref{App:HyperSolidAngle}), and where we have taken
the high temperature limit for the total partition function:
$Z_{\mathrm{tot}}\approx \prod_j [\tilde{\omega}_j/(k_B T)]$.

The condition $Z_\varepsilon/Z_{\mathrm{tot}} <
\alpha_\varepsilon$ then becomes:
\begin{equation}
  T \, > \,\, 2 \, G_D^{1/D} \, \frac{c_\varepsilon^2 \, \tilde{\omega}_{\mathrm{av}}}{k_B \,
  g_{\mathrm{av}}^2} \, \alpha_\varepsilon^{-1/D}
  \,,
\end{equation}
where $g_{\mathrm{av}}=(\prod_n g_n)^{1/D}$ and
$\tilde{\omega}_{\mathrm{av}}=(\prod_n \tilde{\omega}_n)^{1/D}$.

By considering the expression of $c_\varepsilon$ for small
$\varepsilon$ and that of $G_D$ after the Stirling approximation,
and taking into account that for large $D$ one has $(\sqrt{2\pi
D}\,)^{1/D} \rightarrow 1$ and $\alpha_\varepsilon^{-1/D}
\rightarrow 1$, we can just ask:
\begin{equation}
  T \, > \, T_\mathrm{s}(\varepsilon) \equiv  \frac{2^7 \, \mathrm{e} \, M^2\Omega^2 \, \tilde{\omega}_{\mathrm{av}}}{m^2 \, g_{\mathrm{av}}^2 \, k_B\, D} \, \varepsilon^{-1}
  \,, \label{eq:ThresholdOnTheSphere}
\end{equation}
which guarantees that the survival probability of the state
$\Ket{1}$ is always larger than $1-\varepsilon$. In order to
ensure the validity of the previous calculations, condition
(\ref{eq:ThresholdOnTheSphere}) must always be associated to the
high temperature limit condition, $k_B T \gg
\tilde{\omega}_{max}$, which allows the passage to continuum in
Eq.~(\ref{eq:PartFuncSphereEpsilon})

Apart from the technical aspects of the development of the
calculations (on the hypercube and on the hypersphere), the two
threshold temperatures differ because of the regime where they
apply and because of the scaling with the number of oscillators.
In particular, $T_\mathrm{c}(\varepsilon)$ applies when the number
of oscillators is small, and turns out to be proportional to such
number. We do not need such a threshold temperature to be much
higher than the frequencies of the oscillators. On the contrary,
$T_\mathrm{s}(\varepsilon)$ applies in the limit of high
temperature and can be applied even in the case of large number of
oscillators. Such threshold temperature scales as the inverse of
the number of oscillators. This means that by assuming that the
coupling constants scale as $g_k \sim D^{-1/2}$, one obtains that
the threshold temperature $T_\mathrm{s}(\varepsilon)$ is more or
less independent of the number of oscillators.

\begin{figure}
\subfigure[]{\includegraphics[width=0.40\textwidth,
angle=0,clip=a]{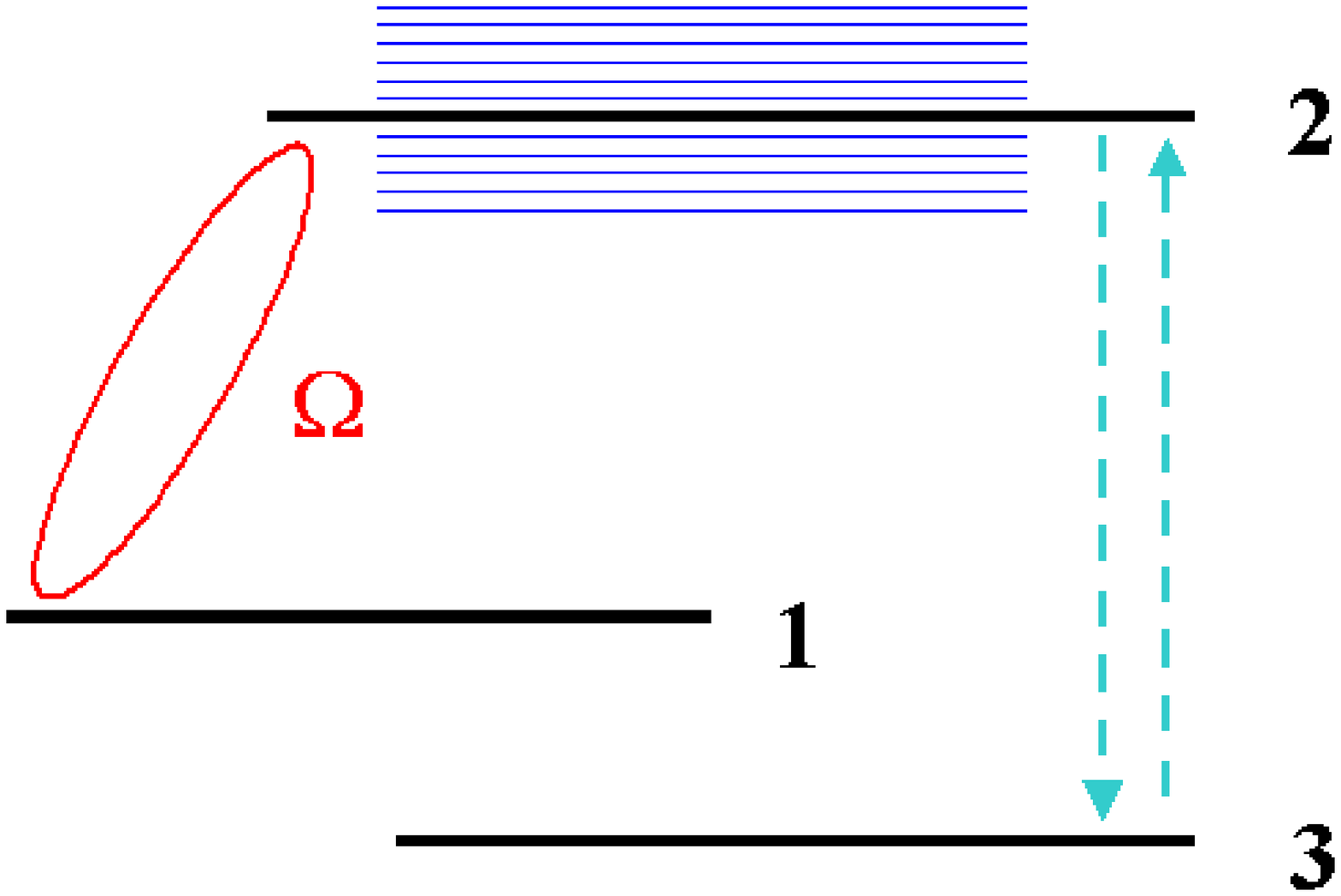}} %
\subfigure[]{\includegraphics[width=0.40\textwidth,
angle=0,clip=b ]{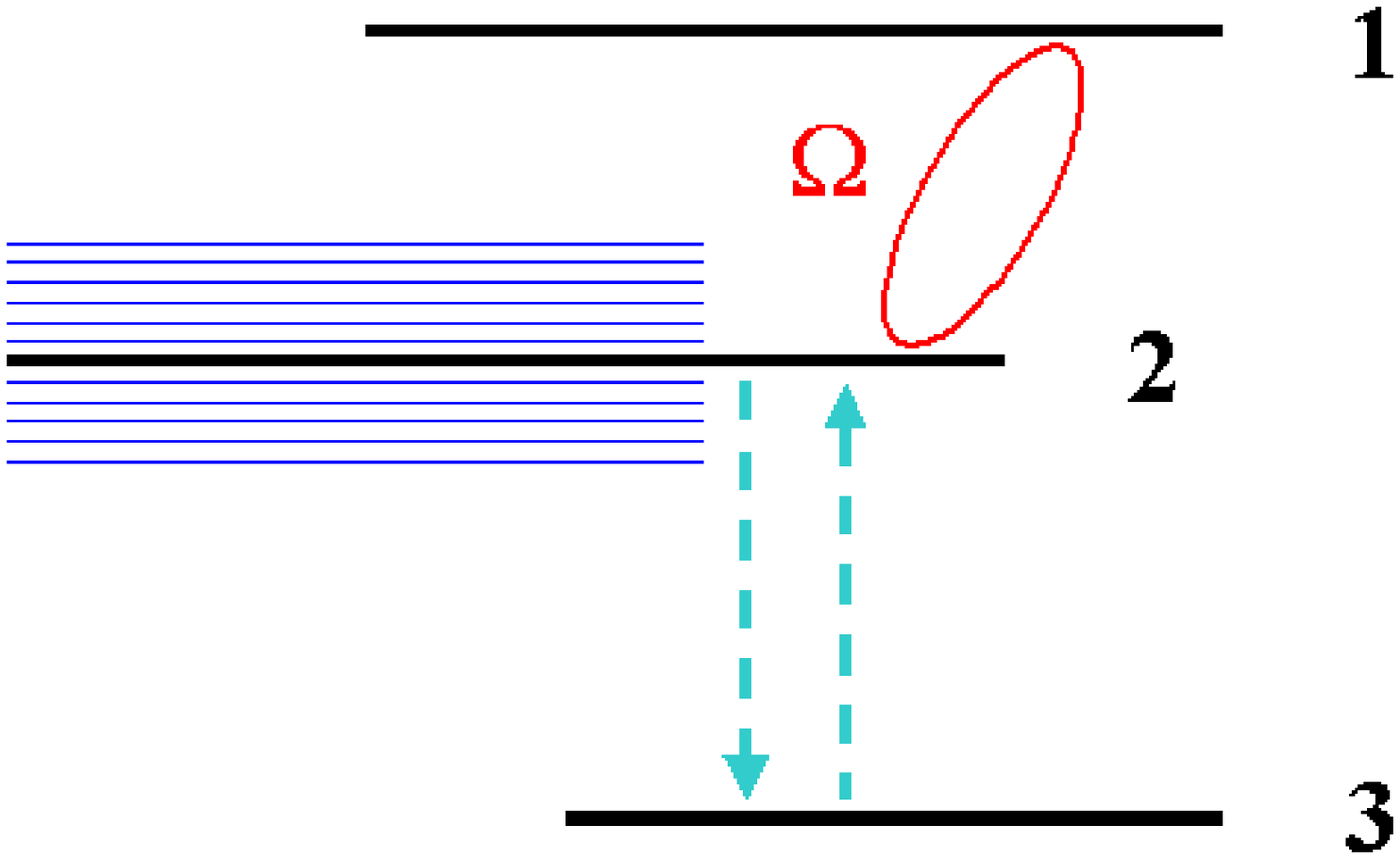}}%
\caption{(Color online). The two possible physical scenarios
corresponding to a bandwidth which does not intersect the Bohr
frequency associated to the $1\rightleftarrows 3$ transitions. In
(a) all the frequencies of the oscillators exceed the frequency
$\omega_1-\omega_3$, while in (b) they are always smaller. The
numerical simulations in Fig.~\ref{fig:Single_Oscillator} and
Fig.~\ref{fig:Many_Oscillators} are both obtained in the scenario
(b).} \label{fig:ReservoirScheme}
\end{figure}

It is worth stressing that, according to the analysis reported in
the appendix \ref{App:Many_Oscillators}, the previous treatment is
valid if one of these two conditions is satisfied: either
$\tilde{\omega}_k > \omega_1-\omega_3$ \, $\forall k$ \, or \,
$\tilde{\omega}_k < \omega_1-\omega_3$ \, $\forall k$, always
requiring $m>0$. This means that we have to avoid those situations
wherein some frequencies of the harmonic oscillators are smaller
than $\omega_1-\omega_3$ while some other are larger. In other
words, the Bohr frequency of the transition $1\rightarrow 3$
should not belong to the frequency band of the oscillators. The
two possible scenarios are pictured in
Fig.~\ref{fig:ReservoirScheme}.

Finally, we emphasize that usually $T_\mathrm{c}(\varepsilon)$
significantly overestimates the temperature necessary to have the
formation of Zeno subspaces. Moreover, both
$T>T_\mathrm{c}(\varepsilon)$ and $T>T_\mathrm{s}(\varepsilon)$
provide only sufficient conditions, so that in principle the Zeno
subspaces can appear for lower temperatures.

\section{Discussion}

In this paper we have analyzed a simple interaction model between
a set of harmonic oscillators initially at thermal equilibrium and
a three-state system. The latter system is prepared in a state
($1$) which is not directly involved in boson-assisted
transitions, but it is coupled to a state ($2$) which undergoes
transitions toward the third state ($3$) due to the interaction
with the bosonic part of the system. Therefore there is \Change{an
effective}{a coupling} between states $2$ and $3$ mediated by the
harmonic oscillators responsible for transitions between such two
states. The states $1$ and $2$ are directly coupled, but when the
\Change{effective coupling between $2$ and $3$}{coupling between
$2$ and $3$ mediated by the oscillators} becomes very strong, a
partitioning of the Hilbert space neutralizes the direct coupling
between $1$ and $2$.

There are two ways to make the \Change{effective}{} coupling
between $2$ and $3$ stronger: one possibility is to consider
higher coupling constants $g_k$'s, which makes the effect occur
even at zero temperature; another possibility is to increase the
temperature of the harmonic oscillators in order to make that
those bosonic states which stronger couple the transitions
$2\rightleftarrows 3$ (due to an high value of $g_k\sqrt{n_k+1}$)
are much involved in the dynamics. The latter case is the one
considered in this paper.

A very remarkable result is the scaling of the threshold
temperature at which the Zeno subspaces surely appear. Indeed, in
the limit of high temperature such threshold temperature scales as
$g_{\mathrm{av}}^{-2} D^{-1}$. Therefore, assuming a larger and
larger number of oscillators, keeping constant the quantity
$g_{\mathrm{av}}^2 D$, we can rigorously assert that even in the
presence of a bosonic bath the partitioning of the Hilbert space
occurs. Instead, if the coupling constants $g_k$'s scale in such a
way that $g_{\mathrm{av}}^2 D$ significantly decreases when $D$
increases, then the temperature necessary to create the Zeno
subspaces becomes higher and higher and could become infinite when
the number of modes of the bosonic field is infinite.

Concerning the limitations of our analysis, it is important to
clarify that the proof given in the appendix
\ref{App:Many_Oscillators} is based on a precise hypothesis about
the diagonal entries of the matrix. Such hypothesis translates
into the assumption that the frequencies of the oscillators are
localized in a finite band, so that it is possible to find a
finite maximum and a non-vanishing minimum --- both of the same
sign --- for the Bohr frequencies related to transitions from
state $1$ to state $3$ assisted by the acquisition of a bosonic
quantum. In other words, the Bohr frequency $\omega_1-\omega_3$
should not be included in the frequency band that contains all the
frequencies of the harmonic oscillators. \Change{This is
physically intuitive, because if any of the oscillators is tuned
very close to the $1\rightarrow 3$ transition, then an effective
coupling between $1$ and $3$ mediated by the oscillators can
raise, avoiding the separation of the state $\Ket{1}$ from the
rest of the Hilbert space.}{} Therefore, the $D\rightarrow\infty$
limit should be considered keeping finite the bandwidth of the
reservoir and ensuring that such band does not intersect the Bohr
frequency $\omega_1-\omega_3$.

As a final remark, we mention that if the subsystem $B$ is made of
spins instead of harmonic oscillators, the previous effect
seemingly does not occur. Indeed, for the partitioning to happen
we need to have $g_k\sqrt{n_k+1}\gg\Omega$, which is fulfilled for
large $n_k$, and then in the majority of the invariant subspaces.
Since such large numbers of excitations are not possible for a
single spin, we guess that it is more difficult to obtain Zeno
subspaces through the interaction with spins at thermal
equilibrium, at least assuming a structure of interaction
analogous to the one we considered in this paper.

\appendix

\section{} \label{App:SurvivalProbability}

Let $\Ket{\psi(0)}$ be the initial state of the physical system,
$\Ket{\phi_k}$ the eigenstates of the Hamiltonian. Introducing
$\BraKet{\psi(0)}{\phi_k} = \sqrt{\chi_k} \,
\mathrm{e}^{i\alpha_k}$, one can write:
\begin{equation}
  \Ket{\psi(0)} = \sum_k \sqrt{\chi_k} \, \mathrm{e}^{-i\alpha_k(0)}\,
  \Ket{\phi_k}\,,
\end{equation}
which evolves into the state
\begin{equation}
  \Ket{\psi(0)} = \sum_k \sqrt{\chi_k} \, \mathrm{e}^{-i\alpha_k(t)}\,
  \Ket{\phi_k}\,.
\end{equation}

If there is a $\chi_{\bar{k}}$, say $\chi_1$, which is larger than
the sum of all the remaining ones, then it is easy to convince
oneself of the following relations:
\begin{eqnarray}
  \nonumber
  |\BraKet{\psi(0)}{\psi(t)}| &=& |\chi_1 + \sum_{k\not=1} \chi_k \, \mathrm{e}^{-i[\alpha_k(t)-\alpha_1(t)]}| \\
  &\ge& |\chi_1 - \sum_{k\not=1} \chi_k | = |2\chi_1-1|\,,
\end{eqnarray}
where we have used $\sum_k \chi_k = 1$ and we have considered the
\lq worst case\rq\, corresponding to $\alpha_k(t)-\alpha_1(t)=\pi$
\, $\forall k\not= 1$.

The condition $\chi_1 > \sum_{k\not=1} \chi_k$ is clearly
equivalent to $\chi_1>1/2$. The survival probability is the
squared modulus of the overlap of the initial state and the
evolved state:
\begin{equation}
  |\BraKet{\psi(0)}{\psi(t)}|^2 \ge (2\chi_1-1)^2\,.
\end{equation}

\section{} \label{App:Many_Oscillators}

Let us consider an Hamiltonian of the following form:
\begin{equation}
  H = \left(
  \begin{array}{ccccc}
    \delta_1 & \Omega & 0 & ... & 0 \cr
    \Omega & \delta_2 & c_1 & ... &  c_D \cr
    0 & c_1 & \delta_3 & ... & 0 \cr
    \vdots  & \vdots & \vdots & \ddots & \vdots \cr
    0 & c_D & 0 & ... & \delta_{D+2} \cr
  \end{array}
  \right)\,,
\end{equation}
in the basis $\Ket{A}$, $\Ket{B}$, $\Ket{C_1}$, ...$\Ket{C_D}$.
Let us assume $\delta_k \not=0 \,\, \forall k$ and introduce the
quantities $c=(\sum_{k=1}^D c_k^2)^{1/2}$,
$m\equiv\min_{k=1}^D(|\delta_{k+2}-\delta_1|)$ and
$M\equiv\max_{k=1}^D(|\delta_{k+2}-\delta_1|)$. For the sake of
simplicity, we are considering the $c_k$'s as real quantities.

In this appendix we prove that in the limit of very large $c$
there is an eigenstate which approaches the state $\Ket{A}$,
provided $ m > 0$. In particular, we will se that there exists an
eigenvalue $\lambda_1$ such that $|\lambda_1-\delta_1| < 2 M
\Omega^2 / c^2$, and that the corresponding eigenstate
$\Ket{\lambda_1}$ has an overlap with $\Ket{A}$ which becomes
closer and closer to unity as $c$ increases.

\begin{widetext}

{\it The secular equation} --- The determinant of the matrix above
can be developed (with respect to the first row) as follows:
\begin{equation}
  \det H = \delta_1\left(
  \begin{array}{cccc}
    \delta_2 & c_1 & ... &  c_D \cr
    c_1 & \delta_3 & ... & 0 \cr
    \vdots & \vdots & \ddots & \vdots \cr
    c_D & 0 & ... & \delta_{D+2} \cr
  \end{array}
  \right)
  -\Omega\left(
  \begin{array}{cccc}
    \Omega & c_1 & ... &  c_D \cr
    0 & \delta_3 & ... & 0 \cr
    \vdots  & \vdots & \ddots & \vdots \cr
    0 &  0 & ... & \delta_{D+2} \cr
  \end{array}
  \right)
  \,,
\end{equation}
where the second minor can be easily evaluated as
$\Omega\prod_{k=3}^{D+2}\delta_k$, while the first minor is
calculated as follows:
\begin{eqnarray}
  \nonumber
  && \delta_2\left(
  \begin{array}{ccccc}
    \delta_3 & 0 & 0 & ... & 0 \cr
    0 & \delta_4 & 0 & ... & 0 \cr
    0 & 0 & \delta_5 & ... & 0 \cr
    \vdots & \vdots & \vdots & \ddots & \vdots \cr
    0 & 0 & 0 & ... & \delta_{D+2} \cr
  \end{array}
  \right)
  -c_1\left(
  \begin{array}{ccccc}
    c_1 & 0 & 0 &... &  0 \cr
    c_2 & \delta_4 & 0 &... & 0 \cr
    c_3 & 0 & \delta_5 &... & 0 \cr
    \vdots  & \vdots & \vdots & \ddots & \vdots \cr
    c_D &  0 & 0 & ... & \delta_{D+2} \cr
  \end{array}
  \right)
  +c_2\left(
  \begin{array}{ccccc}
    c_1 & \delta_3 & 0 &... &  0 \cr
    c_2 & 0 & 0 &... & 0 \cr
    c_3 & 0 & \delta_5 &... & 0 \cr
    \vdots & & \vdots & \ddots & \vdots \cr
    c_D &  0 &  & ... & \delta_{D+2} \cr
  \end{array}
  \right)+...\\
  && \,\,\,\,\,\,\, %
  = \prod_{k=2}^{D+2}\delta_k - c_1^2\prod_{k=4}^{D+2}\delta_k - c_2^2\times\!\!\!\prod_{k=3,5,6...D+2}\delta_k \, + \, ... \,
  =\,
  \left(\prod_{k=2}^{D+2}\delta_k\right)\times \left(1-\sum_{l=1}^D
  \frac{c_l^2}{\delta_2\delta_{l+2}}\right).
\end{eqnarray}
Taking into account these results, one obtains:
\begin{equation}
\det H = \left(\prod_{k=1}^{D+2}\delta_k\right)\times
\left(1-\frac{\Omega^2}{\delta_1\delta_2}-\sum_{l=1}^D
\frac{c_l^2}{\delta_2\delta_{l+2}}\right)\,.
\end{equation}
The eigenvalue equation, $\det (H-\lambda\mathbb{I})=0$, is
obtained through the replacement $\delta_k \rightarrow \delta_k -
\lambda$:
\begin{equation}
P(\lambda) \equiv
\left[\prod_{k=1}^{D+2}(\delta_k-\lambda)\right]\times
\left[1-\frac{\Omega^2}{(\delta_1-\lambda)(\delta_2-\lambda)}-\sum_{l=1}^D
\frac{c_l^2}{(\delta_2-\lambda)(\delta_{l+2}-\lambda)}\right] =
0\,.
\end{equation}
Considering $\lambda=\delta_1+\sigma$, one gets:
\begin{eqnarray}
P(\delta_1+\sigma) &=&
-\sigma\left[\prod_{k=3}^{D+2}(\delta_k-\delta_1-\sigma)\right]\times
\left[\delta_2-\delta_1-\sigma+\frac{\Omega^2}{\sigma}-\sum_{l=1}^D
\frac{c_l^2}{\delta_{l+2}-\delta_1-\sigma}\right]\,=0\,.
\label{eq:PolynomialWithEta}
\end{eqnarray}
\end{widetext}

{\it The eigenvalue close to $\delta_1$} --- Let us first consider
the case where $\delta_{k+2}>\delta_1$ \, $\forall k\ge 1$.

In the limit $\sigma\rightarrow 0$ one obtains:
\begin{equation}\label{eq:PDelta1}
P(\delta_1) = - \Omega^2 \,
\left[\prod_{k=3}^{D+2}(\delta_k-\delta_1)\right]\,,
\end{equation}
from which one immediately finds $P(\delta_1)<0$.

On the other hand, when $c$ is large enough one can prove that
$P(\delta_1+\xi)>0$ for $\xi=2M\Omega^2/c^2$. Indeed, assume $c^2
> 2M\Omega^2/m$ so that $\delta_{l+2}-\delta_1-\xi > 0$ \,
$\forall k\ge 1$. Therefore, since $M > \delta_{l+2}-\delta_1-\xi$
\, $\forall l$ and hence $\sum_l c_l^2/(\delta_{l+2}-\delta_1-\xi)
> \sum_l c_l^2/M = c^2/M$, one obtains:
\begin{equation}
\delta_2-\delta_1-\xi+\frac{\Omega^2}{\xi}-\sum_{l=1}^D
\frac{c_l^2}{\delta_{l+2}-\delta_1-\xi} < \delta_2-\delta_1 -
\frac{c^2}{2M} \,.
\end{equation}
Now, if $\delta_2-\delta_1 < 0$ then we have a negative quantity
for any $c$, and therefore $P(\delta_1+\xi)>0$. If instead
$\delta_2-\delta_1 \ge 0$, we need to take
$c>(2(\delta_2-\delta_1)M)^{1/2}$. Of course we always have to
satisfy the previously considered condition $c^2 > 2M\Omega^2/m$
to guarantee $\delta_{k+2}-\delta_1-\xi > 0$.

Summarizing, for $c>\max\{\Omega\sqrt{2M/m}, \,
[2(\delta_2-\delta_1) M]^{1/2}\}$ and $\xi=2M\Omega^2/c^2$ we are
sure that $P(\delta_1+\xi) > 0$, and we can then assert that for
some $\eta$: $0 < \eta < \xi = 2 M\Omega^2/c^2$ there is an
eigenvalue $\lambda_1=\delta_1+\eta$.

{\it Eigenstate and overlap} ---  If $\Omega\not=0$ and
$\delta_{k+2}-\delta_1-\eta \not= 0$ \, $\forall k$, then one can
easily find that the corresponding eigenstate is:
\begin{eqnarray}
\nonumber \Ket{\delta_1+\eta} &=& {\aleph_\eta}\left[ \Ket{A} +
\frac{\eta}{\Omega}\Ket{B}\right.\\ &-& \left.\sum_{l=1}^D
\frac{\Omega^{-1} c_l\eta}{(\delta_{l+2}-\delta_1-\eta)}\Ket{C_l}
\right]\,,
\end{eqnarray}
which approaches $\Ket{A}$ in the limit of very large $c$. Indeed,
consider first of all that in such limit we have $\eta \rightarrow
0$, since $0 < \eta < \xi = 2 M \Omega^2 / c^2 \rightarrow 0$, so
that the condition $\delta_{l+2}-\delta_1-\eta \not= 0$ is easily
satisfied. Secondly, one finds $|c_l \eta| \, \le \, | 2 M
\Omega^2 c_l/2 c^2 | \, \le \, 2 M \Omega^2 / c \rightarrow 0$,
and on this basis it is immediate to prove that
$|\BraKet{\delta_1+\eta}{A}| \rightarrow 1$.

Now, one cannot be sure that $\delta_{l+2}-\delta_1-\xi > m$, even
for small $\xi$, but assuming $c^2 > 4 M \Omega^2 / m$ one has
$(\delta_{l+2}-\delta_1-\xi)^{-1} < 2/m$ and {\it a fortiori}
$(\delta_{l+2}-\delta_1-\eta)^{-1} < 2/m$. Then, reminding that
$\eta < 2M\Omega^2/c$, one finds:
\begin{eqnarray}
\nonumber
\BraKet{\delta_1+\eta}{A}&=&\left(1+\frac{\eta^2}{\Omega^2}+\sum_{l=1}^D
\frac{\Omega^{-2}\eta^2 c_l^2}{(\delta_{l+2}-\delta_1-\eta)^2}
\right)^{-\frac{1}{2}}\\
\nonumber &\ge& \left(1+\frac{4 M^2 \Omega^2}{c^4}+\sum_{l=1}^D
\frac{16 M^2 \Omega^{2} c_l^2}{m^2 c^4}\right)^{-\frac{1}{2}}\\
\nonumber &=& \left(1+\frac{4 M^2 \Omega^2}{c^4}+\frac{16 M^2 \Omega^{2}}{m^2 c^2}\right)^{-\frac{1}{2}}\\
\nonumber &=& \frac{m c^2}{(m^2 c^4 + 16 M^2 \Omega^2 c^2 + 4 m^2 M^2 \Omega^2)^{1/2}}%
\,,\\
\label{eq:ADelta1Overlap_LowerBound}
\end{eqnarray}
which clearly approaches unity in the limit $c\rightarrow\infty$.
All this reasoning is invalidated if $m=0$.

The condition
\begin{equation}
|\BraKet{\delta_1+\eta}{A}| \ge \sqrt{\chi}\,,
\end{equation}
is satisfied when:
\begin{equation}
c \ge \frac{4\Omega M\,\sqrt{\chi}}{m} \,
\sqrt{\frac{1+\sqrt{1+(1-\chi) m^4/(16 \chi M^2 \Omega^2)} }
{2(1-\chi)}} \,.
\end{equation}

Assuming $m/\Omega$ not diverging and $\chi\rightarrow 1$, we can
neglect the term multiplying $1-\chi$, then requesting just the
following:
\begin{equation}\label{eq:ConditionCChi}
c \, \gtrapprox \, \frac{4 M
\Omega}{m}\,\sqrt{\frac{\chi}{1-\chi}} \,.
\end{equation}

Of course, we also need to have $c>\max\{4M\Omega^2/m, \,
[2(\delta_2-\delta_1) M]^{1/2}\}$ --- which includes the
previously considered $c^2 > 2 M \Omega^2 / m$ ---, but for $\chi$
close enough to unity such condition is surely included in the
condition (\ref{eq:ConditionCChi}).

{\it Generalization and limitations} --- All the previous
discussion keeps its validity if $\delta_{k+2} < \delta_1$ \,
$\forall k \ge 1$, which means that $\prod_{l=3}^{D+2}
(\delta_l-\delta_1)$ appearing in $P(\delta_1)$ --- see
Eq.~(\ref{eq:PDelta1}) --- has the sign $(-1)^D$. Now, define
$\xi'=-2M\Omega^2/c^2$, and assume $c^2>2M\Omega^2/m$ in order to
have $\delta_{k+2}-\delta_1-\xi' < 0$ \, $\forall k$, so that
$\prod_{l=3}^{D+2} (\delta_l-\delta_1-\xi)$ maintains its previous
sign. Since:
\begin{equation}
\delta_2-\delta_1-\xi'+\frac{\Omega^2}{\xi'}-\sum_{l=1}^D
\frac{c_l^2}{\delta_{l+2}-\delta_1-\xi'} > \delta_2-\delta_1 +
\frac{c^2}{2M} \,,
\end{equation}
in order to obtain a change of sign for the quantity
$P(\delta_1+\xi)$ it is sufficient to have:
\begin{equation}
\delta_2-\delta_1 + \frac{c^2}{2M}>0 \,.
\end{equation}

Now, either $\delta_2-\delta_1\ge 0$, so that the quantity on the
right-hand side above is positive whatever the number $c$, or
$\delta_2-\delta_1< 0$, and that quantity can be made positive
provided $c > (2|\delta_2-\delta_1| M)^{1/2}$. This immediately
implies that $P(\delta_1)$ and $P(\delta_1+\xi')$ have opposite
signs, provided $c$ large enough. Therefore, this time we have an
eigenvalue $\lambda_1=\delta_1+\eta'$ with $-2M\Omega^2/c^2 <
\eta' < 0$. The corresponding eigenstate $\Ket{\delta_1+\eta'}$ is
close to $\Ket{A}$, since the inequality we derived for
$\Ket{\delta_1+\eta}$ in (\ref{eq:ADelta1Overlap_LowerBound}) is
valid also for $\Ket{\delta_1+\eta'}$.

Here we are not considering those cases in which $\delta_{l+2} >
\delta_1$ holds for some $l$'s and does not hold for others.

Finally, we remark that if the $c_k$'s are not real ---
corresponding to the interaction
$\sum_k(g_k\hat{a}_k\KetBra{2}{3}+g_k^*\hat{a}_k^\dag\KetBra{3}{2})$,
which generalizes that in Eq.~(\ref{eq:ManyOcillatorsHamiltonian})
--- the previous results keep holding, provided the replacement
$c_k^2 \rightarrow |c_k|^2$ everywhere.

\section{} \label{App:HyperSolidAngle}

The geometric factor $G_D$ in
Eq.~(\ref{eq:PartFuncSphereEpsilonExplicit}) comes from the right
hand-side integral of Eq.~(\ref{eq:PartFuncSphereEpsilon}). In
particular, exploiting to hyperspherical coordinates, such
integral requires integration over the $D-1$ spherical angles. In
general there are $D-2$ angles which can run over $[0,\pi]$ and
one that can run over $[0,2\pi]$. In our particular case, the
spanned intervals are smaller, since the variables $y_j = g_j^2
n_j$ can assume only positive values, and hence all the angles
span $[0, \pi/2]$.

The spherical coordinates are defined by:
\begin{eqnarray}
\nonumber
y_1 &=& r \cos\phi_1, \\
\nonumber
y_2 &=& r \sin\phi_1 \cos\phi_2, \\
\nonumber
y_3 &=& r \sin\phi_1 \sin\phi_2 \cos\phi_3, \\
\nonumber
\vdots \\
\nonumber
y_{D-1} &=& r \sin\phi_1 \sin\phi_2 ... \sin\phi_{D-2}
\cos\phi_{D-1}\\
y_D &=& r \sin\phi_1 \sin\phi_2 ... \sin\phi_{D-2}
\sin\phi_{D-1}\,,
\end{eqnarray}
which correspond to the volume element:
\begin{eqnarray}
  \nonumber
  \prod_k \mathrm{d}y_k &=& r^{D-1} \, \left(\prod_{k=1}^{D-2} \sin^{D-k-1}\phi_k\right) \times \mathrm{d}r \, \prod_{k=1}^{D-1}
  \mathrm{d}\phi_k\,,\\
\end{eqnarray}
that can be straightforwardly derived from the Jacobian of the
coordinate transformation $|\partial y_k / \partial x_j|$, with
$x_j=\phi_j$ for $j\le D-1$ and $x_D=r$.

Therefore one obtains:
\begin{eqnarray}
  \nonumber
  && \!\!\!\!\!\!\!\!\!\!
  \int_{y_n\ge\, 0} ... \int_{\sum_n y_n^2 < c_\varepsilon} \prod_{j=1}^D y_j \mathrm{d}y_j \, \\
  \nonumber
  &=& \, \int_0^{c_\varepsilon} r^{2D-1}\mathrm{d}r \int... \int
  \prod_{k=1}^{D-1} f_k(\phi_k) \, \mathrm{d}\phi_k \, \\
  \nonumber
  &=& \frac{c_\varepsilon^{2D}}{2D} \int_0^{\frac{\pi}{2}}... \int_0^{\frac{\pi}{2}}
  \prod_{k=1}^{D-1} \sin^{2(D-k)-1}\phi_k\cos\phi_k \mathrm{d}\phi_k \\
  &=& \frac{c_\varepsilon^{2D}}{2D} \times \frac{1}{2^{D-1}(D-1)!}  = \frac{c_\varepsilon^{2D}}{2^D D!}   \,.
\end{eqnarray}

On this basis we find that the geometric factor raising from the
integration over the angles is:
\begin{eqnarray}
  G_D = \frac{1}{2^D D!}  \,.
\end{eqnarray}

Exploiting the Stirling approximation~\cite{ref:StirlingApprox},
which is valid for large $D$, one eventually gets:
\begin{eqnarray}
  G_D \approx (2\pi D)^{-1/2} \, 2^{-D} \, D^{-D} \, \mathrm{e}^D \,.
\end{eqnarray}



\begin{thebibliography}{99}

\bibitem{ref:MishraSudarshan} B. Misra and E. C. G. Sudarshan, J. Math. Phys. {\bf 18},  7456 (1997).

\bibitem{ref:Itano1990} W. M. Itano, D. J. Heinzen, J. J. Bollinger and D. J. Wineland, Phys. Rev. A {\bf 41} 2295 (1990).

\bibitem{ref:Fischer1997} S. R. Wilkinson, C. F. Bharucha, M. C. Fischer, K. W. Madison, P. R. Morrow, Qian Niu, Bala Sundaram, and M. G. Raizen, Nature {\bf 387}, 575 (1997).

\bibitem{ref:Schulman1998} L. S. Schulman, Phys. Rev. A {\bf 57}, 1509 (1998).

\bibitem{ref:Panov1999a} A. D. Panov, Phys. Lett. A {\bf 260}, 441 (1999).

\bibitem{ref:Panov1999b} J. Audretsch, M. B. Mensky, A. D. Panov, Phys.Lett. A {\bf 261},
44 (1999).

\bibitem{ref:Militello2011} B. Militello, M. Scala, A. Messina and
N. V. Vitanov, Phys. Scr. T {\bf 143}, 014019 (2011).

\bibitem{ref:Scala2010} M. Scala, B. Militello, A. Messina, and N. V. Vitanov, Phys. Rev. A {\bf 81}, 053847 (2010).

\bibitem{ref:PascazioFacchi2001} P. Facchi and S. Pascazio, {\sl Progress in Optics} {\bf 41}
edited by E. Wolf, Elsevier, Amsterdam, 2001.

\bibitem{ref:Militello2001A} B. Militello, A. Messina and
A. Napoli, J. Phys. A {\bf 286}, 369 (2001).

\bibitem{ref:Militello2001B} B. Militello, A. Messina and
A. Napoli, Fortscr. Phys. {\bf 49}, 1041 (2001).

\bibitem{ref:PascazioFacchi2002} P. Facchi and S. Pascazio, Phys. Rev. Lett. {\bf 89}, 080401
(2002).

\bibitem{ref:PascazioFacchi2004} P. Facchi, D. A. Lidar and S. Pascazio, Phys. Rev. A 69, 032314 (2004).

\bibitem{ref:PascazioFacchi2008} P. Facchi and S. Pascazio, J. Phys. A: Math. Theor.
{\bf 41}, 493001 (2008).

\bibitem{ref:PascazioFacchi2010} P. Facchi, G. Marmo and S. Pascazio, J. Phys.: Conf. Ser.
{\bf 196}, 012017 (2009).

\bibitem{ref:Peres1} A. Peres, Am. J. Phys. {\bf 48}, 931 (1980).

\bibitem{ref:Home1997} D. Home and M. A. B. Whitaker,  Ann. Phys. {\bf 258}, 237 (1997).

\bibitem{ref:App1} Keisuke Fujii and Katsuji Yamamoto, Phys. Rev. A {\bf 82}, 042109
(2010).

\bibitem{ref:App2} Gonzalo Alvarez, D. D. Bhaktavatsala Rao, L. Frydman,
and G. Kurizki, Phys. Rev. Lett. {\bf 105}, 160401 (2010).

\bibitem{ref:App3} R. Sch\"{u}tzhold and G. Gnanapragasam, Phys. Rev. A {\bf 82}, 022120
(2010).

\bibitem{ref:Ruseckas2002} J. Ruseckas, Phys. Rev. A {\bf 66}, 012105 (2002).

\bibitem{ref:Sabrina2006} S. Maniscalco, J. Piilo and K.-A. Suominen,
Phys. Rev. Lett. {\bf 97}, 130402 (2006).

\bibitem{ref:Bhaktavatsala2011} D. D. Bhaktavatsala Rao and Gershon Kurizki,
Phys. Rev. A {\bf 83}, 032105 (2011).

\bibitem{ref:Gardiner} C.~W. Gardiner and P. Zoller, {\it Quantum Noise\/} (Springer-Verlag, Berlin, 2000).

\bibitem{ref:Petruccione} H.-P. Breuer and F. Petruccione, {\it The Theory of Open Quantum Systems\/} (Oxford University Press, Oxford, 2002).

\bibitem{ref:StirlingApprox} M. Abramowitz and I. A. Stegun, {\it Handbook of Mathematical
Functions} (Dover Publications, 1965).


\end{thebibliography}
\end{document}